\documentclass[]{aa}
\usepackage{txfonts}
\usepackage{graphicx}
\usepackage{natbib}
\bibpunct{(}{)}{;}{a}{,}{,}

\begin{document}


\title{Improved distance determination to M51 from supernovae 2011dh and 2005cs}

\author{J. Vink\'o\inst{1,3}\and 
K. Tak\'ats\inst{1}\and 
T. Szalai\inst{1}\and
G. H. Marion\inst{2}\and
J. C. Wheeler\inst{3}\and
K. S\'arneczky\inst{4}\and
P. M. Garnavich\inst{5}\and
J. Kelemen\inst{4}\and
P. Klagyivik\inst{4}\and
A. P\'al\inst{4,6}\and
N. Szalai\inst{4}\and 
K. Vida\inst{4}
}

\institute{Department of Optics and Quantum Electronics, University of Szeged, D\'om t\'er 9, Szeged, 6720 Hungary\\
\email{vinko@physx.u-szeged.hu}\and
Harvard-Smithsonian Center for Astrophysics, 60 Garden St, Cambridge, MA 02138 USA \and
Department of Astronomy, University of Texas at Austin, 1 University Station C1400, Austin, TX 78712-0259, USA \and
Konkoly Observatory of the Hungarian Academy of Sciences, POB 67, Budapest, 1525 Hungary \and
Department of Physics, University of Notre Dame, 225 Nieuwland Science Hall, Notre Dame, IN 46556, USA \and
Department of Astronomy, Lor\'and E\"otv\"os University, POB 32, Budapest, 1518 Hungary
}

\date{Received ; accepted }

\abstract{}
{The appearance of two recent supernovae, SN~2011dh and 2005cs, both in M51,
provides an opportunity to derive an improved distance to their host galaxy
by combining the observations of both SNe.}
{We apply the Expanding Photosphere Method to get the distance to M51 by fitting
the data of these two SNe simultaneously. In order to correct for the
effect of flux dilution, we use correction factors ($\zeta$) appropriate for
standard type II-P SNe atmospheres for 2005cs, but find $\zeta \sim 1$ for 
the type IIb SN~2011dh, which may be due to the reduced H-content of its ejecta.}
{The EPM analysis resulted in $D_\mathrm{M51} = 8.4$ $\pm ~0.7$ Mpc. 
Based on this improved distance, we also re-analyze the $HST$ observations of
the proposed progenitor of SN~2011dh. We confirm that the object detected on
the pre-explosion $HST$-images is unlikely to be a compact stellar cluster. 
In addition, its derived radius ($\sim 277$ $R_\odot$) is too large for being 
the real (exploded) progenitor of SN~2011dh.}
{ The supernova-based distance, $D = 8.4$ Mpc, is
in good agreement with other recent distance estimates to M51. }


\titlerunning{Improved distance to M51}
\authorrunning{Vink\'o et al.}

\maketitle

\section{Introduction}

The recent discovery of supernova (SN) 2011dh \citep{griga} attracted great attention, 
because this was the second bright SN in the nearby galaxy M51 within a few years, 
after  SN~2005cs \citep{kloehr}. Both objects were core-collapse SNe.
SN~2005cs was classified as a peculiar, subluminous, low-velocity type II-P SN
\citep{modjaz05, pastor05csI}, while SN~2011dh turned out to be a
rare type IIb event \citep{arcavi11} with quickly decaying hydrogen 
lines and strengthening He features \citep{marion11a}. Due to the
proximity of the host galaxy ($D \sim 8$ Mpc) and the numerous
pre-explosion observations available, possible progenitors of both SNe have been
successfully detected. SN~2005cs is thought to be originated from a 
$M \sim 8$ $M_\odot$ red supergiant \citep{maund05, li06}, 
while the proposed progenitor of SN~2011dh was a more massive yellow supergiant 
having $M \sim 15$ - 20 $M_\odot$ \citep{maund11, vandyk11}. 
Due to the very early discovery and the quickly responding follow-up
observing campaigns, the time of explosion (actually, the moment of
the initial shock breakout from the envelope of the progenitor object)
has been identified within $\pm 1$ d accuracy for both SNe. 

Having two SNe at the same distance 
with densely sampled light curves and spectra, and accurately
known explosion times offers an unprecedented chance to combine these
datasets and analyze them simultaneously in order to determine an improved
distance to their host galaxy. This is the aim of the present paper. 

The paper is organized as follows. In Sect. 2 we outline the method
we apply for the distance measurement. In Sect. 3 the photometric
and spectroscopic observations are presented, while in Sect. 4
we apply the distance measurement method to these observations.
Finally, we discuss the implications of the improved distance
to the physical properties of the progenitor of SN~2011dh.

\section{Expanding photosphere method}


The Expanding Photosphere Method (EPM) is
a geometric distance measurement method that relates the angular
size of the SN ejecta to its physical radius derived from the observed
expansion velocity of the SN assuming homologous expansion 
\citep{kirshner_epm, dessart05b}. The angular radius of the 
photosphere in a homologously expanding ejecta is defined as
\begin{equation}
\theta = R_\mathrm{phot} / D = {[ R_0 + v_\mathrm{phot} \cdot (t - t_0) ]} / D,
\label{eq-1}
\end{equation}
where $D$ is the distance, $v_\mathrm{phot}$ is the ejecta velocity at the
position of the photosphere, $R_0$ is the radius of the progenitor and
$t_0$ is the moment of the initial shock breakout (usually referred to
as the ,,moment of explosion''). 

In the classical form of EPM the angular radius is derived from
photometric observations via the assumption that the photosphere
radiates as a blackbody, but the opacity is mostly due to electron
scattering, thus the blackbody photons are diluted by scattering 
from free electrons:
\begin{equation}
\theta = {1 \over \zeta(T)} \sqrt{ {f_{\lambda}} \over {\pi B_{\lambda}(T)} },
\label{eq-2}
\end{equation}
where $\zeta (T)$ is the temperature-dependent correction factor describing
the dilution of the radiation, $f_{\lambda}$ is the observed flux and
$B_{\lambda} (T)$ is the Planck function. Expressing the flux in magnitudes,
Eq.\ref{eq-2} can be rewritten as
\begin{equation}
m_{\lambda} = -5 \log{(\theta \zeta(T))} + b_{\lambda}(T)
\label{eq-3}
\end{equation}
where $b_{\lambda}(T)$ is the synthetic magnitude of the blackbody flux 
at temperature $T$ \citep{hamuy99em}.

When the angular radius can be directly measured by high-resolution imaging
(e.g. VLBI), Eq.\ref{eq-1} can be applied directly to get the distance.
This is the ,,Expanding Shock Front Method'' (ESM), which has been 
successfully applied for SN~1993J by \citet{bartel93j}. Very recently,
the expansion of the forward shock of SN~2011dh was detected with 
VLBI \citep{marti11, bietenholz12} and EVLA \citep{krauss12}. 
Although the spatial resolution of these observations was not yet
high enough to use them for distance determination, this method
looks potentially applicable for SN~2011dh in the near future,
when the expanding ejecta reach the necessary diameter.  

In the classical EPM the distance $D$ is inferred via least-squares fitting
to the combination of Eq.\ref{eq-1} and \ref{eq-2}, after neglecting
$R_0$ (which is usually an acceptable approximation 
for epochs greater than $\sim 5$ days after shock breakout):
\begin{equation}
t = D \cdot \left ( \theta \over v_\mathrm{phot} \right ) + t_0
\label{eq-4}
\end{equation}
having $D$ and $t_0$ as the two unknown quantities. In principle, measuring
$\theta$ and $v_\mathrm{phot}$ at two epochs may be sufficient to determine $D$
and $t_0$, in practice at least 5 - 6 observations are needed to 
reduce random and systematic errors. It is important to note that each
observation can provide an independent estimate on $D$ and $t_0$, thus,
the possible errors affecting the results can compensate each other,
provided there are enough data that satisfy the initial assumptions. 

There are several known issues with EPM that must be handled carefully before
applying Eq.\ref{eq-4} to the observations. First, the basic assumption is
that the expansion of the ejecta is spherically symmetric.
It is thought to be valid in most type II SNe, however, there are
cases when asymmetry is observed \citep{leonard04dj}. 
This is usually done via polarization measurements, as electron
scattering may result in detectable net polarization in non-spherical SNe
atmospheres.  
For SN~2005cs, one group presented $R$-band imaging polarimetry \citep{gnedin07}, 
reporting record-high, variable degree of linear polarization (reaching $\sim 8$ \%)
during the early plateau phase. However, the authors do not provide details
on handling the effect of instrumental polarization, and this result has not been
confirmed by other studies so far, so it should be treated with caution. 
For SN~2011dh, there is no such observational indication 
for a non-spherical explosion or expansion yet. Based on the
available pieces of information, we do not find compelling evidence
for asymmetric ejecta geometry in either of these SNe during the
plateau phase. Thus, we assume that both SNe
satisfy the sphericity condition, as it was found earlier 
for most type II SNe. 

Second, Eq.\ref{eq-2} or \ref{eq-3} requires unreddened 
fluxes or magnitudes. Thus, the observed magnitudes/fluxes need to be
dereddened before applying EPM. However, as \citet{E96} and \citet{leonard99em}
pointed out, the distance is only weakly sensitive to reddening 
uncertainties. This is one of the advantages of EPM as a distance
measurement method. 

On the contrary, the results are very sensitive to the values of the 
photospheric velocity. Recently the issue of measuring $v_\mathrm{phot}$ in type II-P
SNe atmospheres was studied in detail by \citet{tv11}. They showed the
applicability of the spectral modeling code SYNOW \citep{branchsynow}
to assign objective and reliable $v_\mathrm{phot}$ to the observed spectra. 
In this paper we use the same technique, i.e. SYNOW to calculate 
$v_\mathrm{phot}$ at each observed phase.  

Finally, the exact values of the correction factors $\zeta (T)$ 
are still debated.
\citet{E96} and \citet{dessart05b} provided $\zeta (T)$ for H-rich 
type II-P atmospheres, but the values from the latter study are systematically 
($\sim$ 50 \%) bigger than from the former one. Moreover, in He-rich 
(type IIb SNe) atmospheres the correction factors are much less known. 
We will discuss this issue in more detail below.

In the followings we exploit the opportunity of having independent datasets 
from two SNe in the same host galaxy M51 (i.e. at the same distance), and
the rare and fortunate circumstance that both of them have known,
accurate ($\pm 1$ d) shock breakout date. If $t_0$ is known independently,
Eq.\ref{eq-4} can be rearranged into the following form:
\begin{equation}
\theta / v_\mathrm{phot} = {1 \over D} \cdot \Delta t,
\label{eq-5}
\end{equation}
where we introduced $\Delta t = t - t_0$, and now $D$ is the only remaining unknown. Combining
the data from both SNe, Eq.\ref{eq-5} can be used to get an improved 
estimate for $D$, which will be applied to SN~2011dh and SN~2005cs in Sect. 4. 
This way we expect to eliminate much of the systematic
uncertainties of EPM (arising from the correlation between the parameters
$t_0$ and $D$) that may sometimes bias the analysis of single SNe.

\section{Observations}

For SN~2005cs extensive data have been published. We have collected
the $BVRI$ light curves from \citet{pastor05csII}. The spectra published
in the same paper were kindly provided to us by Andrea Pastorello. 
  
More recently, SN~2011dh was observed with many instruments, but most of
them are not published at the time of writing this paper.  
In the followings we present some of them that we use in our
analysis. 

\subsection{Photometry of SN~2011dh}

\begin{table}
\centering
\caption{\label{tab:phot} {\it BVRI} photometry of SN 2011dh}
\begin{tabular}{ccccc}
\hline
\hline
JD\tablefootmark{a} & B\tablefootmark{b} & V\tablefootmark{b} & R\tablefootmark{b} & I\tablefootmark{b} \\
\hline
717.5 & 15.158(157) & 14.556(080) & 14.170(128) & ... \\
718.4 & 14.870(114) & 14.274(040) & 13.913(075) & 13.896(075) \\
719.4 & 14.569(123) & 13.949(060) & 13.612(093) & 13.622(098) \\
720.4 & 14.300(090) & 13.676(040) & 13.363(069) & 13.368(075) \\
725.4 & 13.584(057) & 12.889(020) & 12.628(041) & 12.574(048) \\
727.4 & 13.497(051) & 12.770(030) & 12.500(051) & 12.397(058) \\
728.4 & ... & 12.684(060) & 12.399(098) & 12.324(098) \\
729.4 & 13.393(092) & 12.657(050) & 12.369(076) & 12.264(076) \\
730.4 & 13.409(185) & 12.608(100) & 12.303(162) & 12.191(167) \\
731.4 & 13.368(066) & 12.581(020) & 12.279(103) & 12.156(041) \\
732.4 & 13.359(058) & 12.567(030) & 12.252(046) & 12.125(051) \\
733.4 & 13.361(046) & 12.542(030) & 12.205(051) & 12.097(051) \\
734.4 & 13.378(058) & 12.536(030) & 12.180(058) & 12.062(051) \\
735.5 & 13.498(086) & 12.561(050) & 12.202(081) & 12.053(081) \\
738.5 & 13.856(046) & 12.751(030) & 12.305(046) & 12.106(065) \\
739.4 & 14.017(048) & 12.834(020) & 12.371(034) & 12.147(034) \\
740.4 & 14.145(051) & 12.934(030) & 12.444(046) & 12.192(046) \\
742.4 & 14.468(069) & 13.150(040) & 12.566(069) & 12.299(064) \\
748.3 & 15.035(106) & 13.609(040) & 12.900(064) & 12.549(069) \\
749.3 & 15.095(106) & 13.667(040) & 12.944(075) & 12.577(064) \\
751.3 & 15.159(086) & 13.757(050) & 13.038(081) & 12.649(076) \\
752.3 & 15.225(090) & 13.818(040) & 13.073(069) & 12.685(064) \\
754.3 & 15.300(069) & 13.906(040) & 13.160(064) & 12.754(060) \\
755.3 & 15.363(069) & 13.957(040) & 13.196(069) & 12.782(060) \\
\hline
\end{tabular}
\tablefoot{\tablefoottext{a}{JD - 2,455,000.} \tablefoottext{b}{Errors are given in parentheses.}}
\end{table}

For SN~2011dh, ground-based photometric observations were taken from 
Piszk\'estet{\H o} Mountain Station of Konkoly Observatory, Hungary.
We have used the 60/90 cm Schmidt-telescope with the attached $4096 \times 4096$ 
CCD (FoV 70x70 arcmin$^2$, equipped with Bessel $BVRI$ filters). 
In Table~\ref{tab:phot} we present the data for the first 18 nights,
derived by PSF-photometry in
$IRAF$\footnote{IRAF is distributed by the National Optical Astronomy Observatories,
which are operated by the Association of Universities for Research
in Astronomy, Inc., under cooperative agreement with the National
Science Foundation.}. Note that although there are numerous pre-explosion
frames of M51 available, the background at the SN site is relatively
faint and smooth, thus, we did not find the application of image
subtraction necessary to obtain reliable photometric information when
the SN was around maximum light. 

 Transformation to the standard system was computed by applying the
following equations:
\begin{eqnarray}
B - V ~=~ 1.228 \times (b-v) + ZP_{BV} \nonumber \\
V - R ~=~ 0.960 \times (v-r) + ZP_{VR} \nonumber \\
V - I ~=~ 0.934 \times (v-i) + ZP_{VI} \nonumber \\
V ~=~ v + 0.046 \times (V-I) + ZP_{V}.
\end{eqnarray}
The color terms listed  above were obtained
from measurements of Landolt fields on photometric nights. 
The zero-points for 
each night were tied to local tertiary standard stars around M51 collected
from \citet{pastor05csII}.

\begin{figure}
\centering
\resizebox{\hsize}{!}{\includegraphics{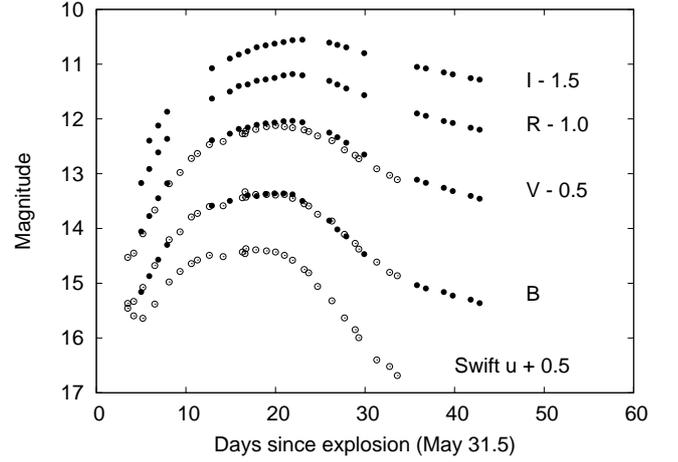}}
\caption{Observed light curve of SN 2011dh. The Konkoly data are plotted with
filled symbols, while $Swift$ $ubv$ data are denoted with open circles. Vertical shifts
are applied between the data corresponding to different filters for better visibility. }
\label{fig-lc}
\end{figure}

In Fig.\ref{fig-lc} we plot and compare our measurements with the $ubv$ data obtained by
$Swift$/UVOT (details for the latter dataset will be given by Marion et al., in prep.).  
The agreement
between the ground-based and space-born optical data (i.e. $B$ and $V$) 
is excellent. We are confident that our photometry for SN~2011dh is
accurate and reliable. 

\subsection{Spectroscopy of SN~2011dh}
\begin{figure}
\centering
\resizebox{\hsize}{!}{\includegraphics{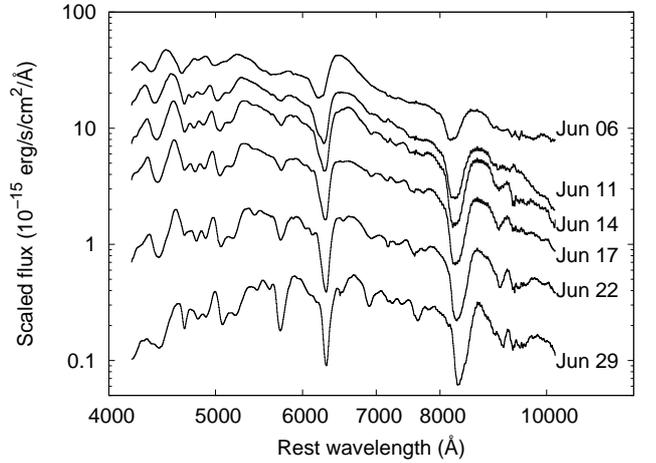}}
\caption{Observed low-resolution HET spectra of SN 2011dh. Note that scales on both axes are logarithmic,
and scaling factors have been applied to separate the spectra for better visibility.}
\label{fig-sp}
\end{figure}

Optical spectra were obtained with the HET Marcario Low Resolution Spectrograph (LRS,
spectral coverage 4200 -- 10200 \AA, resolving power $\lambda / \Delta \lambda$ $\sim 600$) 
at McDonald Observatory, Texas,  on 6 epochs between June 6 and 22, 2011. 
Reduction and calibration of these data were made with standard routines in
$IRAF$. Details will be given in Marion et al. (in prep.). The spectra are shown in 
Fig. \ref{fig-sp}. 

\begin{figure}
\centering
\resizebox{\hsize}{!}{\includegraphics{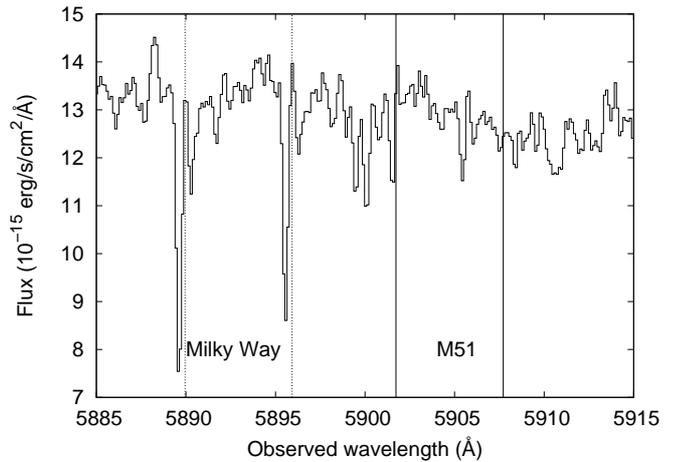}}
\caption{High-resolution HET spectrum of SN 2011dh observed on June 6, 2011
(6 days after explosion).  Vertical lines indicate the
expected positions of the NaD doublet features in the Milky Way (dotted line) 
and in M51 (continuous line). }
\label{fig-hrs}
\end{figure}

In addition, a single high-resolution spectrum were obtained with the HET High-Resolution 
Spectrograph (HRS, spectral coverage 4100 -- 7800 \AA, resolving power $\sim 15,000$)
on June 6, 2011, in parallel with the LRS observations. No narrow features were present
in the spectrum except the Na~D doublet at zero velocity that are expected
to be due to interstellar matter (ISM) in the Milky Way. The spectrum around these
features is plotted in Fig.\ref{fig-hrs}.  We will use the high-resolution
spectrum to constrain the interstellar reddening toward SN~2011dh in Sect. 4.

\section{Analysis}

In the followings we use the data presented in Sect. 3 to estimate the
angular radii and photospheric velocities as a function of phase in order
to perform the EPM-analysis outlined in Sect. 2. 

\subsection{Reddening}

Fortunately, the reddening toward M51 is low. The galactic component,
due to Milky Way ISM, is estimated from the map of \citet{sfd} as 
$E(B-V)_\mathrm{gal} = 0.035$ mag. The equivalent widths of the galactic 
Na~D features shown in Fig.\ref{fig-hrs} are consistent with this
value. As seen, no significant contribution from the M51 ISM is
detected in the SN~2011dh spectrum. Therefore, $E(B-V) = 0.035$ mag
is assumed for SN~2011dh. 

For SN~2005cs we adopted a slightly higher value, $E(B-V) = 0.05$ mag,
after \citet{pastor05csII}. 

\subsection{Photospheric velocities}

\begin{table}
\centering
\caption{\label{tab-epm} Inferred physical parameters for EPM.} 
\begin{tabular}{ccccc}
\hline
\hline
$t - t_0$ & $T_\mathrm{phot}$ & $v_\mathrm{phot}$ & $\theta$ & $\theta / v$ \\
(d) & (K) & (km s$^{-1}$) & ($10^8$ km Mpc$^{-1}$) & (d Mpc$^{-1}$) \\
\hline
\multicolumn{2}{l}{SN~2005cs} &&&  \\
\hline
3.4 & 17\,155 (151) & 7100 (200) & 3.57 (0.05) & 0.58 (0.05) \\  
4.4 & 14\,897 (265) & 6900 (150) & 4.15 (0.08) & 0.69 (0.08) \\
5.4 & 12\,539 (200) & 6500 (275) & 4.85 (0.08) & 0.86 (0.08) \\
8.4 & 12\,530 (127) & 5900 (300) & 4.82 (0.04) & 0.94 (0.04) \\
8.8 & 13\,217 (121) & 5950 (100) & 4.45 (0.04) & 0.86 (0.04) \\
14.4 & 8700 (340) & 4150 (275) & 6.82 (0.01)& 1.90 (0.01) \\
14.4 & 8771 (360) & 4600 (100) & 6.78 (0.05) & 1.67 (0.05) \\
17.3 & 7704 (210) & 4150 (50) & 7.38 (0.04) & 2.05 (0.04) \\
18.4 & 7688 (220) & 4150 (75) & 7.28 (0.11) & 2.03 (0.11) \\
22.4 & 6919 (310) & 3550 (300) &7.81 (0.16) & 2.54 (0.16) \\
34.4 & 6118 (305) & 1900 (50) & 8.68 (0.25) & 5.29 (0.25) \\
36.4 & 5569 (101) & 1800 (100) &  9.05 (0.25) & 6.16 (0.25) \\
44.4 & 5099 (124) & 1350 (50) &  10.02 (0.27)& 8.59 (0.27) \\
61.4 & 4793 (163) & 1050 (50) &  10.43 (0.36)& 11.50 (0.36) \\
\hline
\multicolumn{2}{l}{SN~2011dh} &&& \\
\hline
5.8  & 7624 (94) &11\,400 (200) & 5.26 (0.06)  & 0.53 (0.06) \\
10.8 & 7682 (140)& 9250 (300) & 8.55  (0.29) & 1.07 (0.29) \\
13.8 & 7336 (131) & 7250 (250) & 10.94 (0.41) & 1.74 (0.41) \\
16.8 & 6996 (132) & 7200 (300) & 12.84 (0.45) & 2.06 (0.45) \\
20.8 & 6459 (123) & 7250 (200) & 15.64 (0.57) & 2.49 (0.57) \\
21.8 & 6451 (121) & 7150 (150) & 15.89 (0.61) & 2.57 (0.61) \\
28.7 & 4907 (102) & 5600 (300) & 23.38 (1.19) & 4.83 (1.19) \\
\hline
\end{tabular}
\tablefoot{Errors are given
in parentheses. For the moments of explosion we adopted $t_0$ = 2,453,549.0 JD 
and 2,455,712.5 JD for SNe 2005cs and 2011dh, respectively (see text). 
These epochs were applied to derive the phase of each observation (1st column).}
\end{table}

In this paper we use the photospheric velocities derived from 
fitting SYNOW models to the observed spectra, as described in
\citet{tv11}. As it is shown in that paper, the $v_\mathrm{phot}$ values
determined this way are generally consistent 
(within $\pm$ 10 - 15 \%) with the velocities derived simply from
the absorption minimum of the FeII $\lambda$5169 feature. However,
the FeII velocities may systematically over- or underestimate
the true $v_\mathrm{phot}$. SYNOW has the potential to resolve 
line blending, and thus to give a better estimate of $v_\mathrm{phot}$ than
finding individual (often blended) line minima 
(see \citet{tv11} for details). 

In Table~\ref{tab-epm} the photospheric velocities, as well as
the other physical parameters used for the EPM-analysis are 
presented. 
%

\subsection{Temperatures and angular radii}

The calculation of the angular radii (Eq.\ref{eq-2}) needs observed
fluxes and the effective temperature of the photosphere at the
moments of the photometric observations. We have estimated these
quantities in the following way. First, the observed magnitudes
were dereddened with the $E(B-V)$ values given in Sect. 4.1. Then,
the reddening-free magnitudes were transformed into quasi-monochromatic
fluxes following the calibration given by \citet{bessell98}.  

The photospheric temperatures were estimated by fitting blackbodies
to the spectral energy distributions defined by the contemporaneous
$BVRI$ fluxes. Although this technique is approximate, it has been
shown to produce a good estimate of the effective temperature
for the application of EPM \citep{tv07}. Finally, both the inferred temperatures and
the angular radii have been interpolated to the epochs of the spectroscopic observations. 
These data are summarized in Table~\ref{tab-epm}. 

\subsection{Correction factors for EPM}

As mentioned in Sect. 2, the application of blackbodies to SN 
atmospheres requires correction factors that
take into account electron scattering as well
as line blending and other effects that alter the fluxes
emerging from the photosphere. 
   
As discussed in \citet{tv06}, the angular radii of SN~2005cs 
were calculated by applying the EPM correction factors of \citet{dessart05a}.
However, the data on SN~2011dh were used without corrections, 
i.e. applying $\zeta = 1$. This choice was motivated by 
the IIb classification of SN~2011dh. Because SN~2011dh is
thought to have a He-rich atmosphere  with a thin H-envelope, 
the number of free
electrons in the ejecta should be much less than in type II-P SNe, 
like SN~2005cs. Since the largest contribution to the correction factors
comes from the dilution of blackbody photons that are Thompson-scattered
by free electrons, it is expected that the radiation
of SN~2011dh should be much closer to a blackbody 
than that of SN~2005cs. In other words, due to the lower density of 
free electrons, the photosphere (i.e. the surface of last scattering) 
of SN~2011dh is expected to be much closer to the thermalization depth 
from which the blackbody photons emerge, than in SN~2005cs where the photosphere 
was above the thermalization depth in a H-rich atmosphere. 

This qualitative picture is generally supported by the NLTE models of
the ejecta of SN~1993J by \cite{baron93j}. They modeled the early
spectral evolution of SN~1993J during the first 70 days with various
chemical composition extending from solar He/H ratio to extreme He-rich
atmospheres. For the He-rich compositions they computed substantially
higher correction factors than for solar composition, although with
considerable model-to-model scatter. Since these models may not be
fully applicable to SN~2011dh, we did not attempt to use these correction
factors at their face value, but note that they confirm the qualitative 
expectation that the $\zeta$ values should be higher, i.e. closer to
unity in SN atmospheres having larger He/H ratio.

\subsection{Combination of SNe 2011dh and 2005cs}

\begin{figure}
\centering
\resizebox{\hsize}{!}{\includegraphics{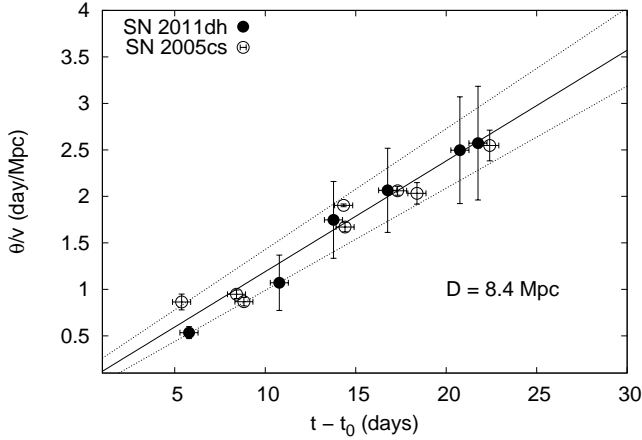}}
\caption{Determination of the distance to M51 in EPM. 
The slope of the fitted (solid) line gives $D^{-1}$ (in Mpc$^{-1}$).
The dotted lines correspond to $1 \sigma$ uncertainty in $D$,
taking into account the random and the systematic errors (see text).}
\label{fig-epm}

\end{figure}

As it was mentioned in Sect. 1, the moment of shock breakout 
($t_0$) is accurately known for both SNe. We have adopted
$t_0$ = 2,453,549.0 JD \citep{pastor05csII} and 
$t_0$ = 2,455,712.5 JD \citep{arcavi11} for SN~2005cs and 
SN~2011dh, respectively. 

We applied Eq.\ref{eq-5} for the data of
both SNe (listed in Table~\ref{tab-epm}) obtained
during the first month, between +5 and +25 days after
explosion. The motivation for using the +5 days lower limit
was that the ejecta may not expand homologously at very
early phases. Also, for epochs earlier than +5 days, 
$R_0$ may not be negligible with respect to the $v_\mathrm{phot} (t - t_0)$
term, as we have assumed to get Eq.\ref{eq-5} (see Sect. 2).  
The restriction of selecting epochs
not later than $\sim 30$ days for EPM was suggested 
by \citet{dessart05b}. They pointed out that the basic 
assumptions of EPM (optically thick ejecta, 
less deviations from LTE condition) are
valid mostly during the first month of the evolution of
the ejecta. Inspecting the data in 
Table~\ref{tab-epm} it can be seen that $\theta / v$ changes
abruptly after $t \sim 22$ days for both SNe, suggesting a 
sudden change in the ejecta physical conditions. 
This epoch is near the time
of maximum light for SN~2011dh. From modeling the light curve of this object, 
Marion et al. (in prep.) pointed out that the early occurence of the peak of the 
light curve could be explained by a sudden drop of the Thompson-scattering 
opacity, possibly due to the recombination of the remaining free electrons and protons. This
is analogous to the condition at the end of the plateau phase in H-rich type II-P 
SNe. The assumptions for EPM are certainly not valid after this phase. 
 
Plotting of $\theta \cdot v^{-1}$ vs $t - t_0$ gives a linear relation
for which the slope equals to $1/D$ (see Eq.\ref{eq-5}). 
Because in this case both SNe are at the same
distance, their data must fall on the same line. 

Fig.\ref{fig-epm} shows that this is indeed the case. Fitting a straight
line to the combined data results in an improved distance to
M51: $$D ~=~ 8.4 ~\rm{Mpc} ~\pm 0.2 ~\rm{Mpc}~\rm{(random)} ~ \pm 0.5 ~\rm{Mpc} ~ \rm{(systematic)}.$$
The systematic error was calculated by taking into account
the $\pm 1$ d uncertainty in $t_0$. 

Our improved distance is in good agreement with the mean distance of M51 listed
in the NED\footnote{http://ned.ipac.caltech.edu} database 
($D \sim 8.03 \pm 0.677$ Mpc), after computing the simple unweighted
average of 15 individual distances. Even though these independent
distance estimates represent an inhomogeneous sample in which each value
may be biased by unknown systematic errors, the $\sim 1 \sigma$ agreement 
between their average and our new value is promising.

\section{Implications for the progenitor of SN~2011dh}

\begin{figure}
\centering
\resizebox{\hsize}{!}{\includegraphics{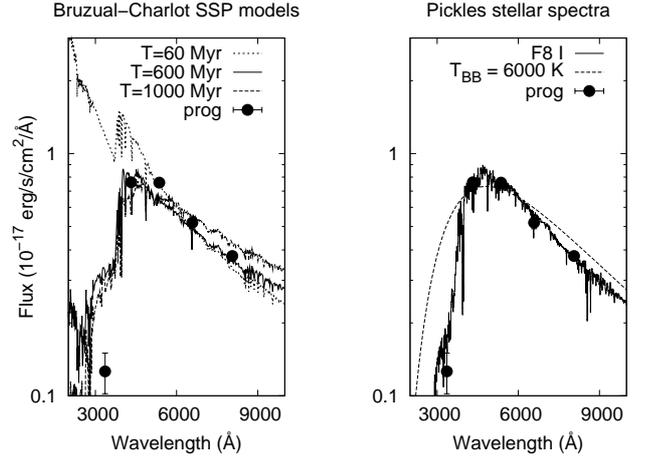}}
\caption{SED of the proposed progenitor (filled circles) and SSP models with cluster ages of 60 Myr, 
600 Myr and 1 Gyr (left), and an F8 supergiant model together with a T=6000 K blackbody (right).
The SED of the 60 Myr-old cluster that may contain a $M$ $>$ 8 $M_\odot$ star is clearly
incompatible with the measured fluxes. The older clusters that fit the data better are
too old to support the presence of such a massive progenitor. }
\label{fig-prog}
\end{figure}

\begin{table}
\centering
\caption{\label{tab-prog} Photometry of the proposed progenitor of SN~2011dh with the
{\it Hubble Space Telescope}}
\begin{tabular}{cccccc}
\hline
\hline
Filter & $\lambda_c^a$ & $m^b$ & $f_{\lambda}^c$ & $M^d$ \\
  & (\AA) & (mag) & (flux) & (mag)\\
\hline 
F336W & 3359 & 23.941(230) & 1.26(24) & $-$5.82(29) \\
F435W & 4318 & 22.405(039) & 7.61(27) & $-$7.32(19) \\
F555W & 5360 & 21.780(030) & 7.59(21) & $-$7.92(19) \\
F658N$^e$ & 6584 & 21.280(064) & 5.19(30) & $-$8.41(20) \\
F814W & 8060 & 21.230(020) & 3.78(07) & $-$8.43(19) \\
\hline
\hline
\end{tabular}
\tablefoot{
\tablefoottext{a}{central wavelength in \AA}\\
\tablefoottext{b}{observed standard magnitude and $1\sigma$ uncertainty}\\
\tablefoottext{c}{dereddened flux (assuming $E(B-V)=0.035$) and its uncertainty in $10^{-18}$ erg s$^{-1}$ cm$^{-2}$ \AA$^{-1}$}\\
\tablefoottext{d}{dereddened absolute magnitude using $D=8.4$ Mpc ($\mu_0 = 29.62$ mag)}\\
\tablefoottext{e}{flight-system magnitudes}}
\end{table}

The improved distance to M51 enables us to revisit the issue of
the progenitor of SN~2011dh identified on pre-explosion
$HST$ frames, immediately after discovery 
\citep{li11, maund11, vandyk11}.

We have analyzed the archival $HST$/ACS and WFPC2 frames
independently from \citet{maund11} and \citet{vandyk11}.
We followed an approach similar to \citet{vandyk11} by
analyzing the individual ,,FLT'' and ,,C0F'' frames obtained
with ACS and WFPC2, respectively, by using the
software packages Dolphot\footnote{{http://purcell.as.arizona.edu/dolphot}}
and HSTphot\footnote{{http://purcell.as.arizona.edu/hstphot}} 
\citep{dolphin00}. We have applied PSF-photometry using
pre-computed PSFs, built-in corrections for charge transfer
efficiency (CTE) losses and the final magnitudes were transformed
into the standard Johnson-Cousins system for the broad-band
filters F336W, F435W, F555W and F814W. These data are collected
in Table~\ref{tab-prog} (the data in the narrow-band F658N filter 
are in the flight-system). Fluxes and absolute magnitudes
are also presented after dereddening with $E(B-V) = 0.035$ mag
(Sect. 4.1), using the magnitude-flux conversion by \citet{bessell98}
and applying the distance modulus $\mu_0 = 29.62$ mag 
corresponding to $D = 8.4$ Mpc (Sect. 4.5). The uncertainties of
the absolute magnitudes reflect the errors of the photometry
and the random plus systematic uncertainty of the distance 
added in quadratures. Generally, these results are in 
between the values presented by \citet{maund11} and 
\citet{vandyk11}, and agree within $1 \sigma$ uncertainty
except for the F336W filter, where our magnitude is 
fainter by $\sim 0.5$ mag ($\sim 2 \sigma$) than that
given by \citet{maund11} and \citet{vandyk11}. 
Given the inferior spatial
resolution of the WFPC2 frames, the faintness of the
source and the issue of the proper source identification
on the WFPC2 images in this crowded region, 
the F336W magnitude in Table~\ref{tab-prog} may only be 
an upper limit. 

In Fig.\ref{fig-prog} left panel, we plot the dereddened 
quasi-monochromatic flux distribution together with model
SEDs of Simple Stellar Populations (SSPs) 
by \citet{bruz03}, assuming cluster ages of
60 Myr, 600 Myr and 1 Gyr, respectively. 
This figure illustrates the conclusion first drawn by
\citet{maund11}, that the SED of the proposed progenitor
is too red to be compatible with a young
stellar cluster hosting a sufficiently massive
star that can become a core-collapse SN. The oldest
cluster that may be able to produce a core-collapse
SN ($M_\mathrm{prog} \sim 8$ $M_\odot$) has the age of 
$\sim 60$ Myr (see e.g. Fig.19 in \citealt{vinko09}).
Fig.\ref{fig-prog} clearly shows that the 60 Myr-old
cluster SED is incompatible with the observed flux
distribution. Those cluster SEDs that are consistent
with the observations are at least an order of
magnitude older, containing only less massive stars.

Strictly speaking, the above argument is valid only
for coeval cluster stars, i.e. when all cluster members
were born by an initial starburst. There is a less
likely, but not unrealistic scenario, in which 
massive stars are formed in an active star-forming region,
and a nearby compact cluster captures them. 
There is observational evidence that young massive clusters
can capture field stars (e.g. Sandage-96 in NGC~2404,
see \citealt{vinko09}), although this process should
occur very (maybe too) fast in order
to explain the presence of a $M > 8$ $M_\odot$ star
in a $t \geq 600$ Myr-old cluster. 

The right panel of Fig.\ref{fig-prog} shows the
good agreement between the observed progenitor
fluxes and the flux spectrum of an F8-type supergiant
having $T_\mathrm{eff} = 6000$ K \citep{maund11, vandyk11}.
Integrating the observed SED, and adopting $T_\mathrm{eff} = 6000$ K
and $D = 8.4 \pm 0.7$ Mpc, the radius of the proposed progenitor 
turns out to be $R_\mathrm{prog} = 1.93 (\pm 0.16) \times 10^{13}$ cm 
(277 $\pm 23$ $R_\odot$). This is very close to the result
of \citet{vandyk11}, who derived $R_\mathrm{prog} \sim 290$ $R_\odot$ from 
the same data, although they adopted a slightly lower distance 
to M51 ($D \sim 7.6$ Mpc).
  
As recently noted by \citet{arcavi11}, \citet{soderberg11} and 
\citet{vandyk11}, the observed early optical-, radio-
and X-ray observations of SN~2011dh are in sharp 
contrast with such a large progenitor radius. All these
pieces of evidence point consistently toward a compact progenitor,
$R_\mathrm{prog} \sim 10^{11}$ cm. Indeed, as discussed by
Marion et al. (in prep.), modeling the bolometric
light curve of SN~2011dh during the first 30 days after explosion
(extending to post-maximum epochs) gives strong support
to the compact progenitor hypothesis. Marion et al. (in prep.)
present an upper limit for the progenitor radius as
$R_\mathrm{prog} \lesssim 2 \times 10^{11}$ cm, in good
agreement with \citet{arcavi11} and \citet{soderberg11}.
The radius of the observed object ($R \sim 277$ $R_\odot$) 
derived above definitely rules out that this object was
the progenitor that exploded. It remains the subject of 
further studies whether such an extended, yellow supergiant 
could be a member of a binary system containing also a hot, compact
primary star, which seems to be currently the most probable
configuration for the progenitor of SN~2011dh, or other
hypotheses, e.g a massive star captured by a compact stellar
cluster outlined above, are necessary to resolve this issue.
Further observations of the SN during the nebular phase 
are also essential in this respect.

\section{Summary}

Based on the results above we draw the following conclusions:

\begin{itemize}
\item{We presented new photometric and spectroscopic observations of
the type IIb SN 2011dh.}
\item{Combining these data with those of the type II-P
SN 2005cs, we derived an improved distance  of $D = 8.4 \pm 0.7$ Mpc to M51 by 
applying the expanding photosphere method.}
\item{Using the updated distance, we reanalyzed archival $HST$ observations
of the proposed progenitor of SN~2011dh. It is confirmed that the object detected
at the SN position is an F8-type yellow supergiant, and it is unlikely to
be the exploded progenitor. }
\end{itemize}

\begin{acknowledgements}

This project has been supported by Hungarian OTKA grant K76816, and
by the European Union together with the European Social Fund through 
the T\'AMOP 4.2.2/B-10/1-2010-0012 grant.
The CfA Supernova Program is supported by NSF Grant AST 09-07903.
JCW's supernova group at UT Austin is supported by NSF Grant
AST 11-09801.
AP has been supported by the ESA grant PECS~98073 and 
by the J\'anos Bolyai Research Scholarship of the Hungarian Academy of 
Sciences. 
KS and KV acknowledges support from the the "Lend\"ulet" Program of 
the Hungarian Academy of Sciences and Hungarian OTKA Grant K-081421.
Thanks are also due to the staff of McDonald and Konkoly Observatories
for the prompt scheduling and helpful assistance during the time-critical
ToO observations. The SIMBAD database at CDS, the NASA ADS and NED 
databases have been used to access data and
references. The availability of these services is gratefully acknowledged.

\end{acknowledgements}

\bibliographystyle{aa} 
\bibliography{sn11dh_v4_aa.bbl} 
\end{document}